\documentclass[sigconf]{acmart}
\usepackage{graphicx}
\usepackage{multirow}
\usepackage{svg}
\usepackage{bm}
\usepackage{amsmath}
\usepackage{amsfonts}
\usepackage{algorithm, algpseudocode}
\AtBeginDocument{%
  }

\copyrightyear{2026}
\acmYear{2026}
\setcopyright{cc}
\setcctype{by}
\acmConference[WWW '26]{Proceedings of the ACM Web Conference 2026}{April 13--17, 2026}{Dubai, United Arab Emirates}
\acmBooktitle{Proceedings of the ACM Web Conference 2026 (WWW '26), April 13--17, 2026, Dubai, United Arab Emirates}
\acmPrice{}
\acmDOI{10.1145/3774904.3792562}
\acmISBN{979-8-4007-2307-0/2026/04}

\begin{document}

\title{Towards Context-aware Reasoning-enhanced Generative Searching in E-commerce}

\author{Zhiding Liu}
\authornote{Work done during an internship at Kuaishou Technology.}
\authornote{Equal contribution.}
\email{zhiding@mail.ustc.edu.cn}
\orcid{0000-0003-0994-473X}
\affiliation{%
  \institution{State Key Laboratory of Cognitive Intelligence, University of Science and Technology of China}
  \city{Hefei}
  \country{China}
}

\author{Ben Chen}
\email{benchen4395@gmail.com}
\authornotemark[2]
\orcid{0000-0003-4495-8686}
\affiliation{%
  \institution{Kuaishou Technology}
  \city{Beijing}
  \country{China}
}

\author{Mingyue Cheng}
\authornote{Corresponding authors.}
\email{mycheng@ustc.edu.cn}
\orcid{0000-0001-9873-7681}
\affiliation{%
  \institution{State Key Laboratory of Cognitive Intelligence, University of Science and Technology of China }
  \city{Hefei}
  \country{China}
}

\author{Enhong Chen}
\email{cheneh@ustc.edu.cn}
\orcid{0000-0002-4835-4102}
\affiliation{%
  \institution{State Key Laboratory of Cognitive Intelligence, University of Science and Technology of China }
  \city{Hefei}
  \country{China}
}

\author{Li Li}
\email{lili0516@mail.ustc.edu.cn}
\orcid{0009-0009-5491-3693}
\affiliation{%
  \institution{State Key Laboratory of Cognitive Intelligence, University of Science and Technology of China }
  \city{Hefei}
  \country{China}
}

\author{Chenyi Lei}
\authornotemark[3]
\email{leichy@mail.ustc.edu.cn	}
\orcid{0000-0001-6287-3673}
\affiliation{%
  \institution{Kuaishou Technology}
  \city{Beijing}
  \country{China}
}

\author{Wenwu Ou}
\email{ouwenweu@gmail.com}
\orcid{0009-0004-2437-6835}
\affiliation{%
  \institution{Kuaishou Technology}
  \city{Beijing}
  \country{China}
}

\author{Han Li}
\email{lihan08@kuaishou.com}
\orcid{0009-0000-9801-9292}
\affiliation{%
  \institution{Kuaishou Technology}
  \city{Beijing}
  \country{China}
}

\author{Kun Gai}
\email{gai.kun@qq.com}
\orcid{0000-0002-3636-3618}
\affiliation{%
  \institution{Kuaishou Technology}
  \city{Beijing}
  \country{China}
}

\renewcommand{\shortauthors}{Zhiding Liu et al.}

\begin{abstract}
Search-based recommendation is one of the most critical application scenarios in e-commerce platforms. Users’ complex search contexts—such as spatiotemporal factors, historical interactions, and current query's information—constitute an essential part of their decision-making, reflecting implicit preferences that complement explicit query terms. Modeling such rich contextual signals and their intricate associations with candidate items remains a key challenge. Although numerous efforts have been devoted to building more effective search methods, existing approaches still show limitations in integrating contextual information, which hinders their ability to fully capture user intent.

To address these challenges, we propose a context-aware reasoning-enhanced generative search framework for better \textbf{understanding the complicated context}. Specifically, the framework first unifies heterogeneous user and item contexts into textual representations or text-based semantic identifiers and aligns them. To overcome the lack of explicit reasoning trajectories, we introduce a self-evolving post-training paradigm that iteratively combines supervised fine-tuning and reinforcement learning to progressively enhance the model’s reasoning capability. In addition, we identify potential biases in existing RL algorithms when applied to search scenarios and present a debiased variant of GRPO to improve ranking performance. Extensive experiments on search log data collected from a real-world e-commerce platform demonstrate that our approach achieves superior performance compared with strong baselines, validating its effectiveness for search-based recommendation.
\end{abstract}




\maketitle

\section{Introduction}
Search-based recommendation is one of the most critical application scenarios of modern e-commerce platforms, enabling users to discover and compare products among millions of offerings efficiently \cite{chen2025onesearch, cheng2022towards, mao2024cross}. The search results directly influence purchase decisions and platform revenue, making its effectiveness crucial for both user satisfaction and business success. Consequently, improving the quality of search-based recommendations has emerged as a key driver of user engagement, conversion rates, and long-term loyalty in online retail ecosystems.

Contextual information constitutes an essential component of search-based recommendation, as it captures users’ implicit preferences that complement explicit query terms, especially when query expressions are noisy or incomplete, which makes it difficult to infer the true underlying intent \cite{cheng2026mind2report}. Rich contextual cues—such as sequential dependencies between adjacent searches (e.g., a user searching for a smartphone followed by a phone case) or the emphasis on particular attributes—can reveal latent connections among target items. Moreover, user profiles, geographic locations, and other behavioral signals provide valuable priors about users’ interest distributions, akin to the collaborative filtering principle of leveraging similar behaviors across user groups. Consequently, effectively extracting and modeling such complex contexts during the search process remains a significant challenge for building more accurate and personalized search-based recommendation systems.

Numerous efforts have been devoted to improving search performance from both matching-based and generative perspectives. The former typically focuses on enhancing query–item relevance through semantic retrieval or embedding models. Representative works include classical lexical-matching methods such as TF-IDF and BM25 \cite{tfidf,bm25}. With the advent of deep learning, subsequent research has proposed neural architectures and tailored training strategies to learn discriminative embeddings for queries and targets \cite{huang2013learning,DPR,khattab2020colbert}. More recently, general-purpose embedding models pretrained on large-scale corpora have been explored \cite{neelakantan2022text,bge}, demonstrating superior performance across diverse retrieval scenarios. In contrast, generative approaches reformulate the search task as a sequence generation problem. Early studies transform items into semantic IDs and model the correlation between a query and its target in an autoregressive manner \cite{DSI,tiger}, with subsequent improvements achieved through diverse aspects like multi-modal information fusion \cite{qarm,cobra} and explicit injection of ranking ability \cite{LTRGR}. Despite these advances, existing solutions still struggle to achieve a \textbf{comprehensive understanding of complex search contexts}, which remains a major obstacle to building more accurate and personalized search-based recommendation systems.


To address these challenges, we propose a context-aware, reasoning-enhanced generative searching framework, namely CRS, which explicitly leverages the world knowledge and multi-step reasoning capabilities of large language models (LLMs) \cite{luo2025time, cheng2025can, guo2025deepseek}, to deeply understand complex search contexts. Concretely, we unify heterogeneous user and item contextual signals into a compact text-based representation or text-semantic identifiers and align them, which allows LLMs to operate directly on a common semantic substrate. Operating on this unified representation, the LLM performs explicit reasoning about latent user intent and its associations with item attributes, recommending candidate items in a generative, end-to-end fashion. To instill stable and effective reasoning behaviors without requiring dense human supervision, we develop a self-evolving post-training paradigm that iteratively alternates supervised fine-tuning (SFT) with reinforcement learning (RL), allowing the model to bootstrap and refine its own reasoning trajectories. Finally, recognizing that standard RL formulations can introduce ranking bias in search scenarios, we introduce a debiased variant of GRPO \cite{grpo} tailored to generative searching to improve ranking fidelity and training stability. Together, these components enable explicit, context-aware reasoning that closes the gap between noisy queries and users’ true intentions, yielding a single end-to-end model that better exploits rich contextual signals for search-based recommendation. Our major contributions are as follows:
\begin{itemize}
    \item We introduce a context-aware, reasoning-enhanced generative searching framework that unifies heterogeneous user and item contexts into a common text-based representation and exploits the world knowledge and reasoning ability of LLMs to explicitly infer users’ latent intent.
    \item We develop a self-evolving post-training paradigm that alternates supervised fine-tuning with reinforcement learning to progressively improve the model’s reasoning capability without requiring dense human supervision.
    \item We identify the ranking bias inherent in standard RL algorithms when applied to search scenarios and propose a debiased GRPO variant tailored for generative searching to enhance ranking fidelity and training stability.
    \item Extensive experiments on real-world e-commerce search logs demonstrate that our approach significantly outperforms strong baselines, validating its effectiveness for search-based recommendation.
\end{itemize}

\section{Related Works}
\subsection{Matching-based Search} %
Deep semantic matching has gradually replaced purely term-based retrieval in modern e-commerce and information retrieval. Traditional lexical matching methods, including TF–IDF weighting and the BM25 ranking function \cite{tfidf, bm25}, compute relevance scores by comparing exact or near-exact term overlaps between queries and documents and have long served as the backbone of large-scale search engines. Building on these foundations, early neural approaches such as the Deep Structured Semantic Model (DSSM) and its subsequent variants \cite{huang2013learning,yu2021dual} employ dual-tower architectures that encode queries and items into a shared embedding space, thereby enabling efficient similarity computation and large-scale retrieval beyond surface term matching. More recently, pre-trained language models have been directly adapted to retrieval \cite{zhao2024dense}. DPR \cite{DPR} learns dense query and document representations and consistently outperforms sparse baselines in open-domain QA and general search. Hybrid or late-interaction models, such as ColBERT and SPLADE \cite{khattab2020colbert, formal2021splade}, further enhance fine-grained token-level matching while retaining scalable retrieval properties. Moreover, recent works explore building general embedding models \cite{neelakantan2022text,bge,qwen3_embedding} pre-trained on a large corpus of sentence pairs with contrastive learning, yielding promising retrieval performance in various domains. Despite these advances, the vast majority of matching-based methods still treat each query largely in isolation and devote limited attention to modeling rich contextual factors, which constrains their ability to capture users’ latent interests and dynamic search intentions.
\subsection{Generative Recommendation and Search}
Generative recommendation has attracted increasing attention as it reformulates recommendation from a ranking problem into an end-to-end text generation task. Tiger \cite{tiger} is among the first to move in this direction by introducing semantic IDs derived from RQ-VAE \cite{rqvae} to represent items and generate sequential recommendations within a standard encoder–decoder architecture. Building on this idea, subsequent studies have proposed various strategies to improve item representations and generation quality, including multi-modal alignment to better integrate heterogeneous content features \cite{qarm, liger, cobra}, collaborative information fusion to leverage user–item interaction signals \cite{colarec, letter, gensar}, and the incorporation of auxiliary learning tasks to enhance generalization and robustness \cite{lc-rec}, all of which have demonstrated promising performance gains.

The generative modeling paradigm has also been extensively explored in the search domain. GENRE \cite{GENRE} first demonstrates the feasibility of using a pretrained T5 model to autoregressively generate target entity titles for retrieval, thereby casting entity search as a sequence generation problem. This line of work is further advanced by DSI and NCI \cite{DSI, NCI}, which introduce a learn-to-encode objective and a prefix-aware weight-adaptive decoder to improve retrieval accuracy and scalability. In parallel, several studies integrate learning-to-rank principles into generative retrieval to strengthen ranking quality. For example, LTRGR \cite{LTRGR} continues training a pretrained generative retriever with a margin-based loss, whereas ListGR \cite{listgr} incorporates supervision derived from the Plackett–Luce model into the generation process. In the e-commerce setting, RARE \cite{liu2025real} introduces the notion of commercial intents to guide generation, and GRAM \cite{gram} designs 16 predefined textual identifier codes to represent items, thereby achieving better alignment with commercial text and background knowledge. More recently, the OneSug \cite{guo2025onesug} and OneSearch \cite{chen2025onesearch} are the first models proposed as the first industrially deployed end-to-end generative framework for large-scale e-commerce search, achieving significant improvements.

Moreover, with the rapid improvement of the reasoning capabilities of large language models (LLMs) \cite{guo2025deepseek}, an emerging line of work has begun to investigate reasoning-enhanced recommendation from both latent and explicit reasoning perspectives. LARES \cite{liu2025lares} and LatentR3 \cite{LatentR3} improve representation capacity by performing additional computations in a latent reasoning space without relying on high-quality chain-of-thought data. By contrast, R$^{2}$ec \cite{you2025text} unifies generative reasoning and discriminative recommendation in a single framework guided purely by reinforcement learning.

\begin{figure*}[htbp]
  \centering
  \includegraphics[width=\linewidth]{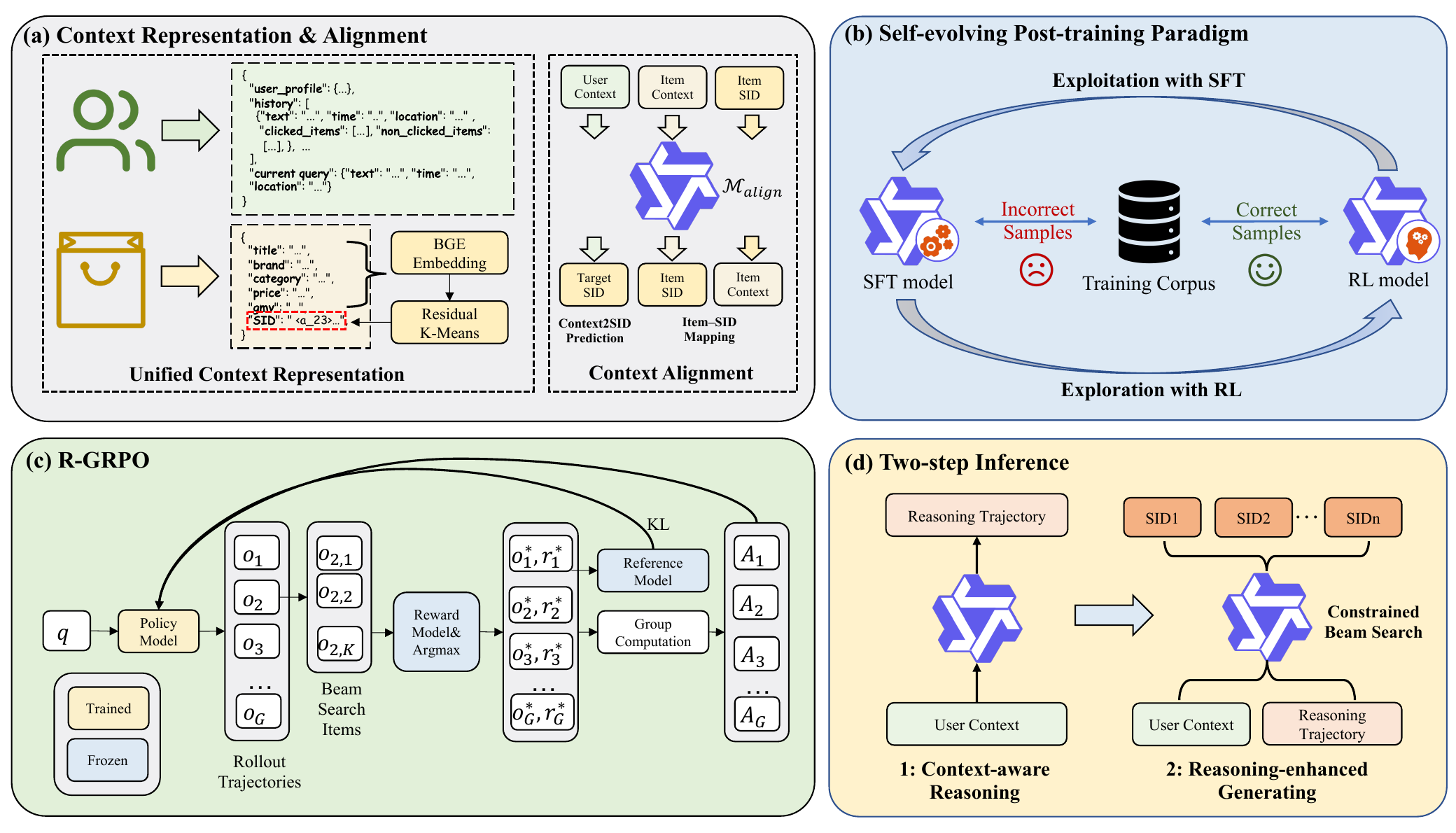}
  \caption{The overview of the proposed CRS framework for search-based recommendation in e-commerce. (a) Context representation and alignment are first conducted, where we unify the complex user and item context into a structured JSON format, and introduce a pre-training stage to align the LLM backbone with e-commerce search context. (b) To tackle the challenge of the scarcity of supervision on the reasoning trajectories, we propose a novel self-evolving post-training paradigm that better stimulates the model's context-aware reasoning ability through alternating between exploration with RL and exploitation with SFT. (c) The debiased RL algorithm, R-GRPO, is proposed for the ranking task, which addresses the optimization target misalignment and reward estimation bias issues. (d) After training, our framework conducts inference in a two-step schema.}
  \label{fig:model}
\end{figure*}

In contrast to prior works, our approach offers a principled and novel perspective by aiming to deeply understand the complex contextual information inherent in e-commerce search. By unifying and aligning heterogeneous user and item context into a shared text-based representation, our framework leverages a slow-thinking–based reasoning mechanism with LLM to more effectively uncover users’ latent intentions and to generate items that better match their true preferences.

\section{Methodology}
In this section, we will delve into the specifics of the proposed CRS framework. As illustrated in Figure~\ref{fig:model}, the framework mainly consists of three key components: context representation and alignment, self-evolving post-training paradigm, and the proposed debiased reinforcement learning algorithm.
\subsection{Context Representation and Alignment}
\subsubsection{Representation}
Unlike conventional approaches that integrate context information through handcrafted features or heterogeneous embedding operations, our framework unifies both user and item contexts into a textual representation, thereby leveraging the world knowledge of LLMs for more robust understanding and reasoning over complex search scenarios. This design alleviates feature-engineering bias and keeps the input format consistent with the pretraining objectives of LLMs.

We argue that contextual signals convey rich implicit preferences, particularly when the current query is noisy or incomplete—a frequent case in search-based recommendation. To capture this, we serialize the user context (including user profile, historical interactions—queries, clicked and non-clicked items—and the metadata of the current query) in chronological order, and embed it directly as structured \texttt{JSON} text without additional feature engineering. Temporal and spatial attributes of both historical and current queries are also retained. A similar process is applied to item context, which incorporates its title, price, brand, category, and gross merchandise volume information. This structured textual form enables LLMs to interpret the meaning of each component more effectively so as to assist the context-aware reasoning process.

Furthermore, to allow the framework to generate target items without heavy post-processing, we convert item contexts into compact semantic IDs (SID) \cite{DSI,tiger}. Specifically, we first transform the JSON texts into initial item embeddings $E^0=\{e_1^0,e_2^0,\cdots,e_{\mathcal{I}}^0\}$ using the bge-small-zh-v1.5 model \cite{bge}, where $\mathcal{I}$ denotes the full item set. We then apply a residual K-Means quantization algorithm \cite{deng2025onerec,qarm} to obtain multi-layer SIDs. Compared to the widely used RQ-VAE approach, this design naturally ensures full codebook coverage without training, avoiding the collapse issue and making it suitable for large-scale SID construction. In detail, the generation process of layer $l$ is expressed as:
\begin{equation}
\begin{aligned}
    C^l &= \{c_1^l,c_2^l,\cdots,c_N^l\} = \text{KMeans}(E^l,N) \\
    s_i^l &= \arg \min_k \|e_i^l-c_k^l\|_2^2, \\
    e_i^{l+1} &= e_i^l - c_{s_i^l}^l, \\
    E^{l+1} & = \{e_1^{l+1},e_2^{l+1},\cdots,e_{\mathcal{I}}^{l+1}\}.
\end{aligned}
\end{equation}
Here $C^l$ is the codebook of the $l$-th layer and $N$ denotes the codebook size. In practice, we employ a codebook of $K=3$ layers augmented with an additional deduplication layer. The codes selected from each layer are concatenated to yield a unique four-token SID for every item (e.g., \verb|<a_23><b_1><c_124><d_0>|). These SIDs of four layers are further registered as \textit{new special tokens} in the LLM’s vocabulary, thereby fully integrating item identifiers into the unified textual representation.

\subsubsection{Alignment}
To bridge the gap between the world knowledge of LLMs and the newly introduced SID tokens, as well as the e-commerce search context, we introduce an additional pre-training stage. This stage consists of two auxiliary tasks, namely \textit{Item–SID Mapping} and \textit{Context2SID Prediction}

\textbf{Item–SID Mapping.} Since SIDs are newly injected into the vocabulary as special tokens representing individual items, we need to ensure that the model must be able not only to retrieve the correct SID given an item’s context but also to reconstruct key attributes of an item from its SID. We therefore introduce two complementary generation directions, \texttt{SID}~$\rightarrow$~\texttt{item\_context} and \texttt{item\_context}~$\rightarrow$~\texttt{SID} to learn the bidirectional mapping between items and their SIDs. This design strengthens the model’s ability to index, memorize and generate items during search, improving retrieval accuracy.

\textbf{Context2SID Prediction.} More importantly, because the textual patterns of search-based recommendation in e-commerce differ markedly from the general pre-training corpus of LLMs, we further instruct the model to directly predict the target item’s SID from the user context without explicit reasoning, i.e., \texttt{user\_context}~$\rightarrow$~\texttt{SID}. This task equips the model with a preliminary ability to perceive and utilize the search context, laying the groundwork for deeper reasoning in later stages.

We apply supervised fine-tuning on these auxiliary tasks to obtain the aligned model \(\mathcal{M}_{\text{align}}\), which serves as a well-initialized starting point for subsequent post-training.

\subsection{Self-evolving Post-training}
Unlike classical reasoning tasks such as mathematics or question answering—where ground-truth reasoning trajectories can be directly elicited from human experts—obtaining sufficiently high-quality traces that faithfully reflect users’ intentions is prohibitively time-consuming and even infeasible, in the e-commerce search setting. This scarcity of explicit supervision constitutes the central obstacle to building a context-aware, reasoning-enhanced model.

A straightforward remedy is to distill trajectories from more powerful LLMs, but we argue that this approach is intrinsically limited by two forms of bias. \textbf{\textit{Data-prior bias}} arises because general-purpose LLMs have little prior knowledge of the vast and dynamic item candidates in e-commerce systems; thus the target item information must be injected into prompts to guide reasoning, which introduces data-leakage and a significant distributional gap between training and inference. \textbf{\textit{Model-capacity bias}} stems from the disparity in reasoning ability between small and large models; simply imitating the reasoning traces of larger LLMs often leads to brittle performance and poor generalization.

To address the aforementioned challenges, we introduce a self-evolving post-training paradigm designed to progressively strengthen the model’s ability to perform deep reasoning over complex search contexts. As illustrated in Figure~\ref{fig:model}(b), the core idea of the paradigm is to alternate between supervised fine-tuning (SFT) and reinforcement learning (RL), coupled with tailored data selection strategies.

Concretely, we initialize from an SFT model $\mathcal{M}_{sft}^0$ obtained by further tuning the aligned model $\mathcal{M}_{align}$ on a small set of context analysis traces distilled from Qwen3-30B-A3B-Instruct-2507 \cite{qwen3technicalreport}. This preliminary distillation step mainly enhances the base model’s instruction-following and slow-thinking ability. We then sample a subset of training instances to efficiently fine-tune the model with RL, thereby encouraging \textbf{\textit{exploration}}. The details of RL training are explained in Section \ref{sec:RL}. Inspired by the filtering strategy of DAPO \cite{yu2025dapo}, we run the model on the training corpus and collect only those cases it gets wrong; training on these “hard” examples incentivizes the model to discover effective policies precisely where its contextual understanding is weak. After obtaining an RL-updated model $\mathcal{M}_{rl}^0$, we perform a subsequent SFT stage to explicitly consolidate the acquired knowledge, i.e., \textbf{\textit{exploitation}}, by training on the correctly solved instances along with their generated reasoning trajectories. This yields a refined SFT model $\mathcal{M}_{sft}^1$, thus completing one iteration of the self-evolving loop. The training procedure consists of several iterations. 

Formally, iteration $i$ can be expressed as:
\begin{equation}
\begin{aligned}
\mathcal{D}_{rl}^i &= \text{Sample}(\text{Incorrect}(\mathcal{D}_{train}, \mathcal{M}_{sft}^i)),\\
\mathcal{M}_{rl}^i &= \text{RL}(\mathcal{D}_{rl}^i, \mathcal{M}_{sft}^i),\\
\mathcal{D}_{sft}^i &= \text{Correct}(\mathcal{D}_{train}, \mathcal{M}_{rl}^i),\\
\mathcal{M}_{sft}^{i+1} &= \text{SFT}(\mathcal{D}_{sft}^i, \mathcal{M}_{rl}^i),\\
\end{aligned}
\end{equation}
where $\mathcal{D}_{train}$ is the full training corpus, $\mathcal{D}_{rl}^i, \mathcal{D}_{sft}^i$ denote the RL and SFT training subsets at iteration $i$ respectively.

\subsection{Debiased Reinforcement Learning}\label{sec:RL}
Reinforcement learning constitutes the core exploration mechanism of our framework, enabling the model to discover effective reasoning strategies over complex contexts. In this section, we first detail our reward design, then discuss how conventional RL algorithms can become biased in search-based recommendation settings, and present our debiased algorithm R-GRPO.

\subsubsection{Reward Design}
To optimize the proposed CRS framework with RL, we design fine-grained, multi-objective reward functions aimed at simultaneously enhancing recommendation accuracy and improving reasoning behavior. These rewards operate at two levels: (1) the reasoning format (syntactic validity and completeness), and (2) the task outcome (prediction accuracy and validity).

\textbf{Reasoning Format Rewards}. To ensure that generated reasoning paths are syntactically valid and sufficiently detailed, we introduce two complementary rewards:
\begin{itemize}
    \item \textit{Reasoning Structure Reward}. We assign a binary reward to indicate whether the model correctly follows the required context-reasoning–before–item structure:
    \begin{equation}
        \mathrm{r}_{structure} = 
        \begin{cases}
        0 & \text{if } \texttt{<think>}, \texttt{</think>} \text{ tags exist}, \\
        -1 & \text{otherwise}.
        \end{cases}
    \end{equation}
    \item \textit{Reasoning Length Reward}. Motivated by recent evidence that test-time scaling can improve reasoning quality, we positively reward more comprehensive reasoning trajectories, while capping their length to avoid overthinking:
    \begin{equation}
    \mathrm{r}_{length} = \text{min}(1, \frac{\text{len(ans)}}{L}),
    \end{equation}
    where $L$ denotes the maximum permitted length.
\end{itemize}

\textbf{Task-outcome Rewards}. To ensure the generated items match users' interests, we introduce two task-related rewards to assess the final predictions' quality.
\begin{itemize}
    \item \textit{SID Accuracy Reward}. Leveraging the hierarchical structure of items’ semantic IDs, we introduce a partial-matching reward rather than a strict binary signal. This yields denser feedback and greater flexibility in measuring similarity between predicted and target items:
    \begin{equation}
    \mathrm{r}_{sid\_acc} = \sum_{i=1}^K w_i \cdot \mathrm{I}(\text{SID}_{pred}^i=\text{SID}_{target}^i).
    \end{equation}
    Here $w_i$s are predefined hyperparameters for weighting the SID accuracy of different layers.
    \item \textit{SID Validity Reward}. Since the generation process can produce invalid SIDs absent from the item set, we introduce a validity check:
    \begin{equation}
    \mathrm{r}_{sid\_val} = 
    \begin{cases}
    1 & \text{if SID}_{pred} \text{ exists in item set}, \\
    0 & \text{otherwise}.
    \end{cases}
    \end{equation}
\end{itemize}

The total reward used for RL training is the sum of the above components.

\subsubsection{Debiased Algorithm: R-GRPO}
Recent years have seen rapid progress in reinforcement learning (RL), and many algorithms have achieved impressive results on challenging reasoning tasks. Nevertheless, we argue that simply transplanting these advances into our search-based recommendation scenario can lead to suboptimal outcomes due to two inherent drawbacks.

First is the optimization target misalignment. Search-based recommendation is fundamentally a \textit{ranking} problem, where the quality of the final ranked item list directly determines user satisfaction. Yet most existing RL methods in generative retrieval follow a rollout–optimize paradigm: a context-aware reasoning trajectory is generated together with a single item prediction in each rollout, and the optimization then maximizes the likelihood (or reward) of the top-1 predicted item. This training objective overlooks the quality of the full ranked list and provides no explicit incentive for better ranking discrimination. Moreover, such a paradigm exacerbates a deeper issue of reward estimation. Under our rule-based reward scheme, algorithms such as GRPO may assign the same reward to two trajectories whose top-1 predictions are both incorrect, even if one trajectory ranks the true target much higher than the other. This conflation ignores valuable information about partial correctness and ranking quality, ultimately limiting the model’s ability to learn nuanced ranking behaviors.

To overcome the above challenges, we develop a debiased reinforcement learning algorithm on top of GRPO that is specifically tailored to search-based recommendation. As shown in Figure~\ref{fig:model}(c), the proposed method, named R-GRPO, redesigns both the rollout procedure and the reward computation so as to explicitly incorporate ranking information into policy optimization. 

Concretely, we decouple the traditional rollout into two stages: (i) generation of reasoning trajectories and (ii) beam search of predicted items. Given a context $q$, the model first generates $G$ context-aware reasoning trajectories $\{o_i\}_{i=1}^G$. For each trajectory, we then conduct an independent beam search to obtain a ranked list of $K$ candidate items, resulting in a total of $G\cdot K$ rollouts $\{o_i^j\}_{i=1,j=1}^{G,K}$. Each rollout is assigned a scalar reward $\{r_i^j\}_{i=1,j=1}^{G,K}$ according to our reward system. To more effectively integrate ranking information into RL training, for each reasoning trajectory, we select the item prediction with the highest reward as the optimal outcome and compute a rank-aware, weighted-sum reward to reflect the entire ranked list. That is,
\begin{equation}
\begin{aligned}
    o_i^* &= o_i^{\text{argmax}(\{r_i^j\}_{j=1}^K)},\\
    r_i^* &= \frac{1}{W}\sum_{n=1}^K\frac{r_i^n}{1+\log n},\\
    W &= \sum_{n=1}^K\frac{1}{1+\log n}.
\end{aligned}
\end{equation}
This design encourages the model not only to produce a single correct item but also to generate reasoning trajectories that lead to higher-quality ranked lists overall.

Finally, the optimization is conducted under the standard GRPO objective as follows:
\begin{equation}
\begin{aligned}
\mathcal{J}_{\text{R-GRPO}}(\theta) = 
&\mathbb{E}_{\substack{q \sim P(Q), \\ \{o_{i}^*\}_{i=1}^{G} \sim \text{argmax }\mathcal{M}_{{\text{old}}}(o|q)}} 
\Bigg\{ 
\frac{1}{G}\sum_{i=1}^{G}  
\frac{1}{|o_i^*|} 
\bigg\{
\sum_{t=1}^{|o_i^*|} \\
&
\min\bigg[ 
r_t(\mathcal{M})\cdot A_i,
\text{clip}\Big(r_t(\mathcal{M}), 1-\epsilon, 1+\epsilon \Big) A_i 
\bigg] \\
&
- \beta \mathbb{D}_{KL} [\mathcal{M} || \mathcal{M}_{ref}] 
\bigg\} 
\Bigg\},\\
r_t(\mathcal{M}) &= \frac{\mathcal{M}(o_{i,t}^*|q, o_{i,<t})}{\mathcal{M}_{\text{old}}(o_{i,t}^*|q, o_{i,<t})}, \\
A_i &= \frac{r_i^*-mean(\{r_1^*,r_2^*,\dots,r_G^*\})}{std(\{r_1^*,r_2^*,\dots,r_G^*\})}.
\end{aligned}  
\end{equation}

Note that we are focusing on modifying the rollout process and the corresponding reward estimation schema to inject the ranking information, therefore our approach can be easily extended to diverse variants of recent RL algorithms \cite{yu2025dapo, gspo}. We summarize the complete training workflow of R-GRPO in Section \ref{sec:alg} to better clarify the procedure of the training process.

\subsection{Inference}
At inference time, we employ the final model obtained after post-training to generate target items conditioned on the given user context. To balance efficiency and accuracy, we adopt a two-step inference strategy rather than directly applying beam search over both reasoning tokens and SID tokens. In the first step, the model produces a deep reasoning trajectory that reflects its internal understanding of the input user context. In the second step, we concatenate the original user context with the generated reasoning trajectory and perform beam search to predict the target SIDs, which is constrained by a pre-computed prefix trie \cite{trie,GENRE} to ensure that only valid output sequences are produced.

\section{Experiments}
\subsection{Experimental Setup}
\subsubsection{Datasets} Though there exist some open-source search-based recommendation datasets \cite{sun2023kuaisar, liu2023jdsearch}, they lack the essential non-anonymous contexts required by our approach. Therefore, to evaluate the effectiveness of the proposed CRS framework, we curated three datasets from the log data of Kuaishou's e-commerce platform. Each dataset contains rich and complex search contexts, including user profile features, spatial-temporal information, historical interactions (including the click and non-click interactions), and metadata of the current query. Specifically, based on the user's target item, we randomly sampled 100K and 50K users across all product categories, and additionally collected 27K users from the fashion category. For every user, we retained the 10 most recent interactions to capture short-term behavioral context. This process yields three benchmark datasets, denoted as All-100K, All-50K, and Fashion-27K, respectively. The detailed statistics of these datasets are summarized as follows in Table \ref{tab:dataset}.
\begin{table}[htbp]
  \centering
  \caption{Dataset statistics}
    \begin{tabular}{l|rrr}
    \toprule
    \textcolor[rgb]{ .2,  .2,  .2}{Dataset} & \multicolumn{1}{l}{\textcolor[rgb]{ .2,  .2,  .2}{All-100K}} & \multicolumn{1}{l}{\textcolor[rgb]{ .2,  .2,  .2}{All-50K}} & \multicolumn{1}{l}{\textcolor[rgb]{ .2,  .2,  .2}{Fashion-27K}} \\
    \midrule
    \textcolor[rgb]{ .2,  .2,  .2}{Num. Users} & \textcolor[rgb]{ .2,  .2,  .2}{99,683} & \textcolor[rgb]{ .2,  .2,  .2}{49.967} & \textcolor[rgb]{ .2,  .2,  .2}{27,219} \\
    \textcolor[rgb]{ .2,  .2,  .2}{Num. Items} & \textcolor[rgb]{ .2,  .2,  .2}{1,072,791} & \textcolor[rgb]{ .2,  .2,  .2}{632,196} & \textcolor[rgb]{ .2,  .2,  .2}{323,481} \\
    \textcolor[rgb]{ .2,  .2,  .2}{Num. Unique Query} & \textcolor[rgb]{ .2,  .2,  .2}{362,859} & \textcolor[rgb]{ .2,  .2,  .2}{202,436} & \textcolor[rgb]{ .2,  .2,  .2}{103,812} \\
    \textcolor[rgb]{ .2,  .2,  .2}{Avg. Query Per User} & \textcolor[rgb]{ .2,  .2,  .2}{6.55} & \textcolor[rgb]{ .2,  .2,  .2}{6.54} & \textcolor[rgb]{ .2,  .2,  .2}{6.77} \\
    \textcolor[rgb]{ .2,  .2,  .2}{Avg. Click Per User} & \textcolor[rgb]{ .2,  .2,  .2}{1.81} & \textcolor[rgb]{ .2,  .2,  .2}{1.81} & \textcolor[rgb]{ .2,  .2,  .2}{1.47} \\
    \bottomrule
    \end{tabular}%
  \label{tab:dataset}%
\end{table}%

\begin{table*}[htbp]
  \centering
  \caption{Performance comparison on the three benchmark datasets, where Qwen3-E is short for Qwen3-Embedding. The best results are marked \textbf{bold} and the second best are marked \underline{underline}.}
    \begin{tabular}{c|c|cccc|cccc|cc}
    \toprule
    \multirow{2}[4]{*}{Dataset} & \multirow{2}[4]{*}{Metric} & \multicolumn{4}{c|}{Matching-based} & \multicolumn{4}{c|}{Generative} & \multicolumn{2}{c}{Ours} \\
\cmidrule{3-12}          &       & TF-IDF & BM25  & BGE   & Qwen3-E & GENRE & DSI   & LTRGR & LatentR3 & CRS   & Imp. (\%) \\
    \midrule
    \multirow{5}[2]{*}{All-100K} & HR@1  & 0.0319  & 0.0411  & 0.0482  & 0.0469  & 0.0728  & 0.0736  & 0.0702  & \underline{0.0891}  & \textbf{0.0919 } & 3.14  \\
          & HR@5  & 0.1004  & 0.1353  & 0.1524  & 0.1605  & 0.1656  & 0.1737  & 0.1683  & \underline{0.1955}  & \textbf{0.2054 } & 5.06  \\
          & HR@10 & 0.1505  & 0.1930  & 0.2301  & 0.2329  & 0.2133  & 0.2303  & 0.2229  & \underline{0.2419}  & \textbf{0.2570 } & 6.24  \\
          & NDCG@5 & 0.0666  & 0.0885  & 0.1005  & 0.0996  & 0.1194  & 0.1255  & 0.1194  & \underline{0.1463}  & \textbf{0.1511 } & 3.28  \\
          & NDCG@10 & 0.0827  & 0.1071  & 0.1256  & 0.1244  & 0.1364  & 0.1438  & 0.1372  & \underline{0.1612}  & \textbf{0.1678 } & 4.09  \\
    \midrule
    \multirow{5}[2]{*}{All-50K} & HR@1  & 0.0412  & 0.0548  & 0.0560  & 0.0550  & 0.0658  & 0.0656  & 0.0676  & \underline{0.0810}  & \textbf{0.0880 } & 8.64  \\
          & HR@5  & 0.1303  & 0.1654  & \underline{0.1767}  & 0.1715  & 0.1423  & 0.1624  & 0.1651  & 0.1743  & \textbf{0.1953 } & 10.53  \\
          & HR@10 & 0.1935  & 0.2297  & 0.2330  & \underline{0.2355}  & 0.1917  & 0.2197  & 0.2182  & 0.2243  & \textbf{0.2498 } & 6.07  \\
          & NDCG@5 & 0.0867  & 0.1113  & 0.1225  & 0.1189  & 0.1056  & 0.1155  & 0.1185  & \underline{0.1298}  & \textbf{0.1441 } & 11.02  \\
          & NDCG@10 & 0.1071  & 0.1322  & 0.1436  & 0.1451  & 0.1215  & 0.1340  & 0.1317  & \underline{0.1460}  & \textbf{0.1617 } & 10.75  \\
    \midrule
    \multirow{5}[2]{*}{Fashion-27K} & HR@1  & 0.0272  & 0.0312  & 0.0345  & 0.0374  & 0.0371  & 0.0437  & 0.0389  & \underline{0.0485}  & \textbf{0.0713 } & 47.01  \\
          & HR@5  & 0.0709  & 0.0826  & 0.0841  & 0.0815  & 0.0881  & \underline{0.0936}  & 0.0871  & 0.0856  & \textbf{0.1359 } & 45.19  \\
          & HR@10 & 0.0981  & 0.1153  & 0.1120  & 0.1201  & 0.1171  & \underline{0.1223}  & 0.1190  & 0.1058  & \textbf{0.1690 } & 38.18  \\
          & NDCG@5 & 0.0495  & 0.0571  & 0.0601  & 0.0609  & 0.0633  & \underline{0.0695}  & 0.0639  & 0.0680  & \textbf{0.1040 } & 49.64  \\
          & NDCG@10 & 0.0581  & 0.0677  & 0.0692  & 0.0732  & 0.0727  & \underline{0.0787}  & 0.0741  & 0.0744  & \textbf{0.1152 } & 46.38  \\
    \bottomrule
    \end{tabular}%
  \label{tab:main_results}%
\end{table*}%

\subsubsection{Baselines}
To comprehensively assess the effectiveness of the proposed CRS framework in search-based recommendation, we compare it against a diverse set of baselines covering both traditional match-based retrieval and recent generative search models:
\begin{itemize}
    \item \textbf{TF-IDF} and \textbf{BM25} \cite{tfidf,bm25} are classical lexical-matching methods that measure similarity based on term overlap.
    \item \textbf{BGE} \cite{bge} and \textbf{Qwen3-Embedding} \cite{qwen3_embedding} are unified embedding models pretrained on large-scale corpora and used to retrieve candidate items via embedding similarity. Here we use the bge-small-zh-v1.5 and Qwen3-Embedding-0.6B model for embedding computing.
    \item \textbf{GENRE} and \textbf{DSI} \cite{GENRE,DSI} are representative generative retrieval models that directly generate target items in an end-to-end manner using a pretrained T5 backbone \cite{xue2021mt5}.
    \item \textbf{LTRGR} \cite{LTRGR} extends the generative paradigm with a learning-to-rank principle to enhance ranking performance.
    \item \textbf{LatentR3} \cite{LatentR3} adopts a latent reasoning paradigm to further improve generative recommendation quality.
\end{itemize}

\subsubsection{Experiments Details and Evaluation Protocols}

To ensure a fair comparison across all generative models, we adopt a unified item codebook configuration. Specifically, the number of codebook layers $K$ is fixed to 3, with layer sizes set to 512, 256, and 256 for the three datasets, respectively. For the LLM backbones, we employ mT5-base for GENRE, DSI, and LTRGR, and Qwen3-0.6B for LatentR3 and our proposed CRS framework, following the configurations recommended in their original implementations. All models are trained until convergence over multiple epochs. In addition, consistent with CRS, LatentR3 incorporates user context as input, whereas the remaining methods rely solely on the current query text. \textit{Notably, we observe comparable performance for GENRE, DSI, and LTRGR when their backbones are replaced with Qwen3-0.6B}.



For evaluation, we split the datasets by user to generate the training and evaluation sets. The same prefix trie of our framework is utilized to guide the beam search process of all the generative models to guarantee the fairness of evaluation. We employ two widely used top-$N$ metrics: NDCG@$N$ (Normalized Discounted Cumulative Gain) and HR@$N$ (Hit Rate). $N$ is set to 1, 5, 10 for a comprehensive evaluation on both retrieval and ranking performance. All models are implemented in PyTorch and optimized using the Adam optimizer.

\subsection{Main Results}

Table~\ref{tab:main_results} summarizes the overall performance and relative improvements of our framework compared with a broad set of baselines. As shown, generative approaches consistently outperform matching-based methods across all datasets and evaluation metrics. We attribute this advantage to the autoregressive modeling paradigm of generative models, which captures richer query–item interactions than simple embedding similarity. Furthermore, the introduction of structured item identifiers (SIDs) allows semantically related items to share common SID prefixes, effectively preserving structural and semantic proximity and leading to additional performance gains. Among the baselines, LatentR3 achieves the strongest performance by incorporating rich contextual information as input, highlighting the importance of explicit context modeling in search-based recommendation.


More importantly, our proposed CRS framework consistently delivers superior performance across all three datasets. These results demonstrate the effectiveness of CRS in deeply understanding user search contexts and inferring user intent through explicit reasoning. In particular, compared with the strongest baseline, CRS achieves relative improvements of \textbf{3–11\%} on the large-scale datasets (All-100K and All-50K) and up to \textbf{47\%} on the more challenging Fashion-27K dataset. Unlike the other two datasets, Fashion-27K contains a large number of highly similar items within the same category, making them difficult to distinguish based on simple query–item correlations. The substantial gains observed on this dataset further validate the strength of our approach, especially in scenarios that require fine-grained intent understanding and reasoning.




\subsection{Ablation Study on R-GRPO}
Beyond the self-evolving post-training paradigm, the proposed debiased reinforcement learning algorithm (R-GRPO) also plays a critical role in enhancing the context-aware reasoning ability of the CRS framework. To validate its effectiveness, we conduct an ablation study on the All-50K dataset, comparing the performance of models trained with R-GRPO against those trained with the baselineGRPO. The results, shown in Table \ref{tab:r-grpo}, indicate that R-GRPO consistently outperforms GRPO across all three RL training iterations. These findings confirm our expectation and provide strong evidence that R-GRPO effectively mitigates the inherent drawbacks in conventional RL algorithms for search-based recommendation.

\begin{table}[htbp]
  \centering
  \caption{Performance comparison between R-GRPO and GRPO. N@K refers to NDCG@K.}
    \begin{tabular}{cc|ccccc}
    \toprule
    \multicolumn{2}{c|}{Model} & HR@1  & HR@5  & HR@10 & N@5 & N@10 \\
    \midrule
    \multirow{3}[2]{*}{$\mathcal{M}_{rl}^0$} & GRPO  & 0.0746  & 0.1631  & 0.2136  & 0.1208  & 0.1379  \\
          & R-GRPO & 0.0766  & 0.1657  & 0.2154  & 0.1229  & 0.1390  \\
          & Imp.(\%) & 2.68  & 1.59  & 0.84  & 1.74  & 0.80  \\
    \midrule
    \multirow{3}[2]{*}{$\mathcal{M}_{rl}^1$} & GRPO  & 0.0828  & 0.1753  & 0.2313  & 0.1310  & 0.1491  \\
          & R-GRPO & 0.0838  & 0.1773  & 0.2336  & 0.1327  & 0.1508  \\
          & Imp.(\%) & 1.21  & 1.14  & 0.99  & 1.30  & 1.14  \\
    \midrule
    \multirow{3}[2]{*}{$\mathcal{M}_{rl}^2$} & GRPO  & 0.0851  & 0.1845  & 0.2391  & 0.1364  & 0.1539  \\
          & R-GRPO & 0.0863  & 0.1873  & 0.2424  & 0.1390  & 0.1568  \\
          & Imp.(\%) & 1.41  & 1.52  & 1.38  & 1.91  & 1.88  \\
    \bottomrule
    \end{tabular}%
  \label{tab:r-grpo}%
  \vspace{-0.1in}
\end{table}%

\subsection{Generalizability on Insufficient Context}

\begin{figure}[htbp]
  \centering
  \includegraphics[width=\linewidth]{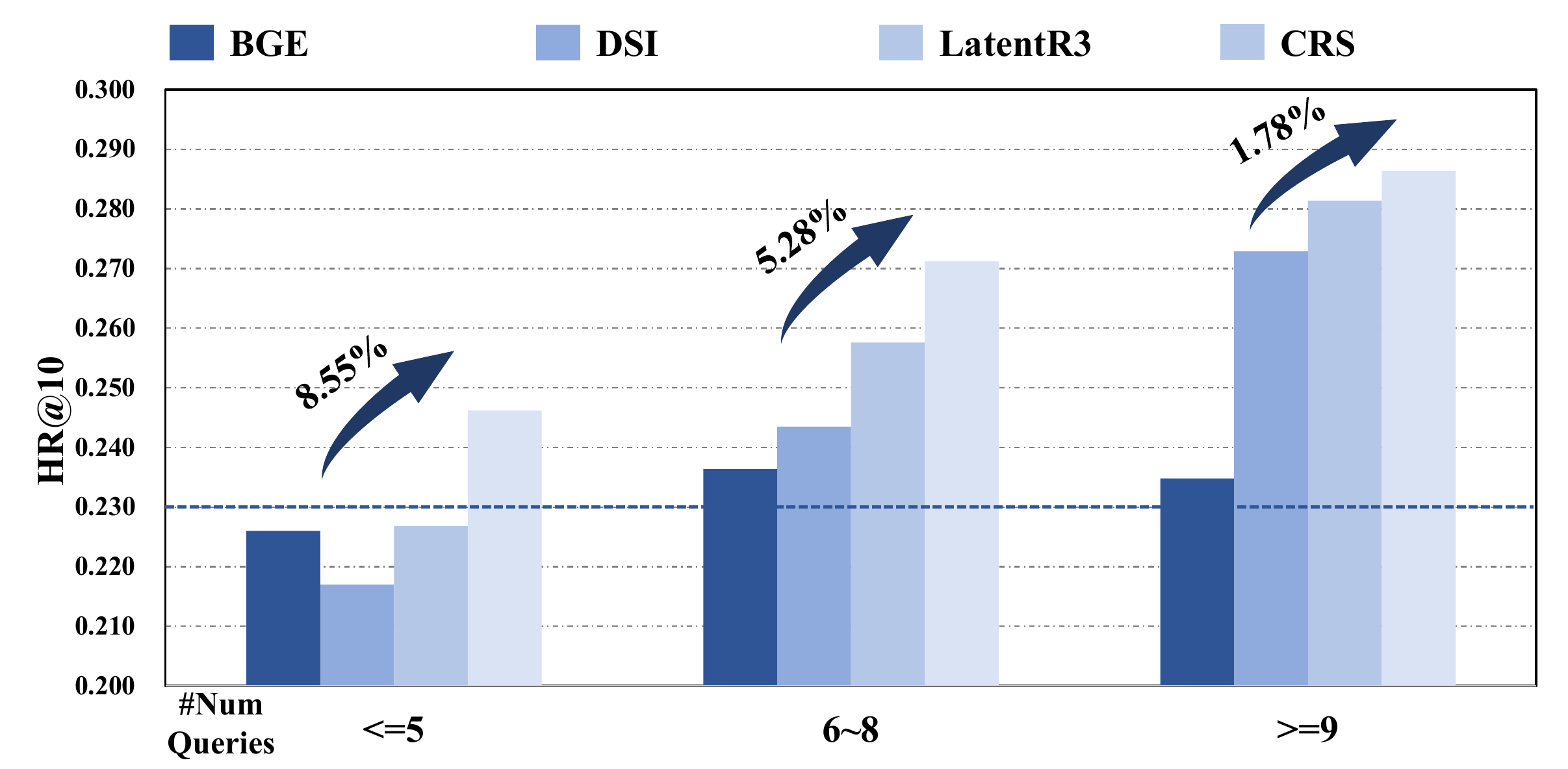}
  \caption{HR@10 evaluation on the three constructed user groups comparing CRS with representative match-based and generative methods.}
  \label{fig:cold_start}
  \vspace{-0.1in}
\end{figure}

In this section, we further investigate the generalization capability of the proposed CRS framework under scenarios with insufficient user context. Specifically, users in the All-100K dataset are partitioned into three groups based on the number of their historical queries, representing varying levels of contextual availability ranging from cold-start to context-rich users. We compare CRS with representative matching-based and generative baselines, including BGE, DSI, and LatentR3, to examine how explicit context-aware reasoning benefits users with limited interaction history. The results are illustrated in Figure~\ref{fig:cold_start}.

As shown in the figure, the matching-based BGE model exhibits relatively stable performance across all user groups, suggesting that it is largely insensitive to the amount of available user context and primarily relies on query–item semantic similarity. This behavior also reflects the semantic homogeneity of user queries in the dataset. In contrast, LatentR3 consistently outperforms DSI across all groups, highlighting the importance of incorporating richer contextual signals for modeling user intent in e-commerce search scenarios.

Furthermore, the proposed CRS framework consistently achieves substantial improvements over all baseline models for users at different context levels. Notably, CRS demonstrates a pronounced advantage in cold-start settings, achieving a relative performance gain of \textbf{8.55\%} for users with minimal historical interactions. This observation indicates that the explicit reasoning mechanism in CRS enables more effective inference of user intent and latent preferences by leveraging structured reasoning over limited contextual evidence, rather than relying solely on surface-level query–item correlations. Overall, these results validate the robustness and practical applicability of CRS, particularly in real-world scenarios where user context is often sparse or incomplete.

\subsection{Scalability Analysis}

To further evaluate the scalability of the proposed CRS framework, we increase the backbone model size to larger configurations with 1.7B and 4B parameters and report the corresponding results on the Fashion-27K dataset in Table~\ref{tab:scale}. As the model capacity grows, the reasoning capability of the backbone is correspondingly enhanced, leading to consistent performance improvements across all evaluation metrics. These results demonstrate that CRS effectively benefits from larger backbone models and validate the strong scalability of the proposed framework.

\begin{table}[htbp]
  \centering
  \caption{Performance of CRS with different backbone sizes on the Fashion-27K dataset.}
    \begin{tabular}{c|ccccc}
    \toprule
    Size & HR@1  & HR@5  & HR@10 & NDCG@5 & NDCG@10 \\
    \midrule
    0.6B  & 0.0713  & 0.1359  & 0.1690  & 0.1040  & 0.1152  \\
    \midrule
    1.7B  & 0.0737  & 0.1370  & 0.1716  & 0.1051  & 0.1172  \\
    Imp.(\%) & 3.37  & 0.81  & 1.54  & 1.06  & 1.74  \\
    \midrule
    4B    & 0.0752  & 0.1460  & 0.1766  & 0.1107  & 0.1214  \\
    Imp.(\%) & 5.47  & 7.43  & 4.50  & 6.44  & 5.38  \\
    \bottomrule
    \end{tabular}%
  \label{tab:scale}%
  \vspace{-0.15in}
\end{table}%


\section{Conclusion}
In this paper, we proposed a novel context-aware reasoning framework for search-based recommendation in e-commerce, namely CRS. Unlike traditional matching-based or generative methods, CRS unified complex user and item information into a structured representation through context representation and alignment, enabling the model to understand the e-commerce search context in a more systematic way. To address the scarcity of supervision signals for reasoning, we designed a self-evolving post-training paradigm that alternates between exploration with reinforcement learning and exploitation with supervised fine-tuning, progressively enhancing the model’s context reasoning ability. Furthermore, we introduced a debiased reinforcement learning algorithm tailored for ranking-oriented optimization, effectively mitigating reward bias and target misalignment issues. Comprehensive experiments on multiple large-scale e-commerce datasets demonstrate the superiority of CRS over both matching-based and generative baselines, yielding consistent and substantial gains across all evaluation metrics.

\begin{acks}
This research was supported by grants from the National Natural Science Foundation of China (No. 62502486, No. U23A20319), the grants of Provincial Natural Science Foundation of Anhui Province (No. 2408085QF193). This research was also funded by Kuaishou Research Project.
\end{acks}

\bibliographystyle{ACM-Reference-Format}
\balance
\bibliography{sample-base}

\appendix
\begin{figure*}[ht]
  \centering
  \includegraphics[width=\linewidth]{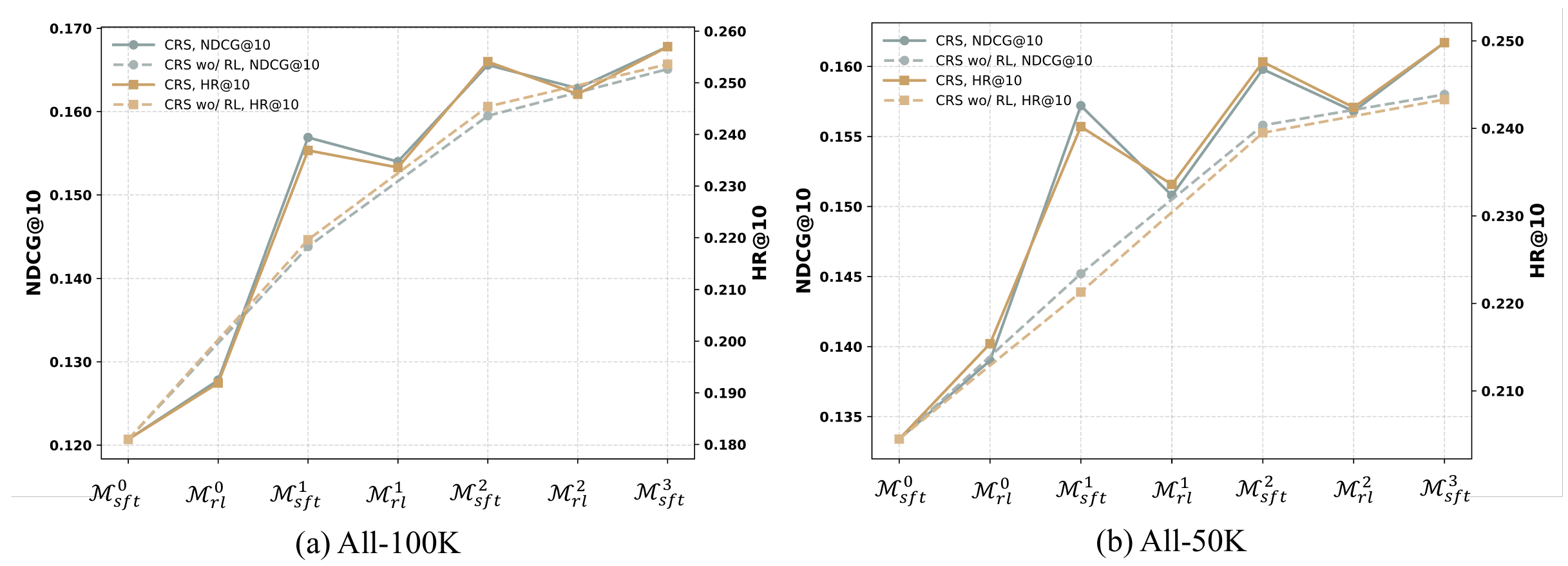}
  \caption{Performance evolution of the proposed CRS framework during self-evolving post-training on the All-100K and All-50K datasets, measured by NDCG@10 and HR@10.}
  \label{fig:self_evolving}
\end{figure*}
\section{Algorithm of R-GRPO}\label{sec:alg}
To better clarify the procedure of the training process, we summarize the complete training workflow of R-GRPO in Algorithm \ref{alg:rgrpo}. Compared to the original GRPO algorithm, our proposal mainly modifies the rollout process and the corresponding reward estimation schema. By explicitly introducing the ranking information, the R-GRPO algorithm focuses on modeling the potential ranking ability of each reasoning trajectory and therefore leads to performance gain for search-based recommendation. Moreover, since the algorithm is independent from the policy optimization procedure, it can be easily extended to diverse variants of recent RL algorithms.

\begin{algorithm}[h]
\caption{R-GRPO: Debiased Reinforcement Learning for Generative Search}
\label{alg:rgrpo}
\begin{algorithmic}[1]
\Require 
Training query set $\mathcal{Q}$; \\
Policy model $M_{\theta}$; reference model $M_{\text{ref}}$; \\
Number of reasoning rollouts $G$; beam width $K$; \\
Clipping parameter $\epsilon$; KL coefficient $\beta$
\ForAll{$q \in \mathcal{Q}$}
    \State \textbf{// Stage 1: Reasoning Trajectory Generation}
    \State Rollout $G$ reasoning trajectories $\{o_i\}_{i=1}^{G} \sim M_{\theta}(\cdot \mid q)$
    
    \State \textbf{// Stage 2: Item Beam Search and Reward Assignment}
    \For{$i = 1$ to $G$}
        \State Perform beam search conditioned on $(q, o_i)$
        \State Obtain ranked item list $\{o_i^{(j)}\}_{j=1}^{K}$
        \For{$j = 1$ to $K$}
            \State Compute reward $r_i^{j}$ using Eq.~(3)--(6)
        \EndFor
        \State Select optimal rollout $o_i^{*} = \arg\max_j r_i^{(j)}$
        \State Compute rank-aware aggregated reward:
        \[
        r_i^{*} = \frac{1}{W} \sum_{n=1}^{K} \frac{r_i^{n}}{1 + \log n},
        \quad
        W = \sum_{n=1}^{K} \frac{1}{1 + \log n}
        \]
    \EndFor
    
    \State \textbf{// Policy Optimization}
    \State Update policy parameters $\theta$ using standard GRPO by maximizing Eq.~(8)
\EndFor
\end{algorithmic}
\end{algorithm}

\section{Investigation of Self-evolving Post-training}




To assess the effectiveness of the proposed self-evolving post-training paradigm, we present in Figure~\ref{fig:self_evolving} the model performance across successive training iterations, evaluated using NDCG@10 and HR@10 on the All-100K and All-50K datasets.

As shown in the figure, iterative post-training substantially and consistently enhances the performance of the overall framework. In particular, compared with the initially distilled model $\mathcal{M}_{\text{sft}}^0$, CRS achieves relative improvements of up to \textbf{39.02\%} and \textbf{21.21\%} in terms of NDCG@10 on All-100K and All-50K, respectively. These significant gains strongly validate our hypothesis that a simple distillation approach may struggle to generalize due to data–prior and model–capacity biases. In contrast, the proposed self-evolving paradigm enables the model to progressively refine its reasoning behavior, thereby strengthening its ability to perform deep and context-aware reasoning over complex search scenarios.

We further observe that RL updates in some iterations may introduce temporary performance degradations. To better understand this phenomenon, we compare CRS with a variant trained using only iterative SFT. This variant consistently underperforms the full CRS framework, highlighting the critical role of reinforcement learning, which expands the policy space through exploration and ultimately leads to stronger and more robust performance.

\section{Reasoning Trajectories Showcases}
We present a comprehensive input-output example with some steps omitted in Table \ref{tab:reasoning case} to better illustrate how the CRS framework reason over complex search context and infer the hidden intents of the user. Note that we also observe potential issues of repetition or inconsistency in some reasoning trajectories, which is one of the limitation of the current design.

\begin{table*}[h]
\centering
\vspace{-0.05in}
\caption{Case study of an input-output example.} 
\vspace{-0.05in}
\label{tab:reasoning case}
\renewcommand{\arraystretch}{1}
\begin{tabular}{@{} p{\dimexpr\textwidth-4\tabcolsep} @{}}
\toprule
\textbf{Instruction:}\\
The following provides a brief user profile and historical search records of an e-commerce user. Please analyze the user's click preferences when searching for items from both the user profile analysis and historical search record analysis perspectives, and based on this, predict the SID of the target product.\\
\midrule
\textbf{Input:}\\
\{\\
\quad "\textbf{user profile}":\{"age":"31-40", "resident area":"northwest, Xining", ...\},\\
\quad "\textbf{history}":[\\
\quad \quad...,\\
\quad \quad \{"text":"iphonewatch strap", "time":"06/10 13:04:30", "location":"Xining", "clicked\_items":[], "non\_clicked\_items":\\
\quad \quad \quad [\{"title":"Compatible with Apple Watch Series 10/9/8/7 Braided Magnetic Buckle Ultra2 Premium-feel SE Stylish Watch Band",\\
\quad \quad \quad "brand":"none", "category":"Smart Device Accessories", "price":"28.8", "gmv":"58.2", "SID":"\texttt{<a\_253><b\_490><c\_253><d\_0>}"\}, \\ \quad \quad \quad ...]\\
\quad \quad \},\\
\quad \quad ..., \\
\quad \quad \{"text":"badminton bag", "time":"06/15 18:28:58", "location":"Xining", "clicked\_items":\\
\quad \quad \quad [\{"title":"Yonex Badminton Backpack Dual-Shoulder Professional Portable Women's Model YY 3-Piece Set Large\\
\quad \quad \quad Capacity Backpack Men's Model Shoe Compartment", "brand":"YONEX", "category":"Badminton Bag", "price":"189.0",\\
\quad \quad \quad "gmv":"3041.0", "SID":"\texttt{<a\_156><b\_105><c\_95><d\_0>}"\},\\
\quad \quad \quad ...],\\
\quad \quad "non\_clicked\_items":\\
\quad \quad \quad [\{"title":"Waterproof Badminton Racket Bag New Large Capacity Sports Backpack Unisex College Student Backpack \\
\quad \quad \quad Dual-Shoulder Bag", "brand":"none", "category":"Badminton Bag", "price":"39.99", "gmv":"604.92",\\
\quad \quad \quad "SID":"\texttt{<a\_158><b\_98><c\_139><d\_0>}"\},\\
\quad \quad \quad ...]\\
\quad \quad\},\\
\quad \quad \{"text":"Badminton uniform", "time":"06/15 18:36:34", "location":"Xining", "clicked\_items":\\
\quad \quad \quad [\{"title":"2024 New Yonex Badminton Apparel Blue and White Porcelain YY Tournament Uniform Quick-Dry Breathable Sports\\
\quad \quad \quad Short-Sleeve for Men and Women", "brand":"YONEX", "category":"Badminton Uniform", "price":"35.0", "gmv":"1400.0",\\
\quad \quad \quad "SID":"\texttt{<a\_156><b\_95><c\_95><d\_0>}"\},\\
\quad \quad \quad ...],\\
\quad \quad "non\_clicked\_items":\\
\quad \quad \quad [\{"title":"Li Zijia Competition-Grade Short-Sleeve Set Badminton New-Style Quick-Dry Breathable Sports Competition Training\\
\quad \quad \quad Uniform", "brand":"none", "category":"Badminton Uniform", "price":"88.00", "gmv":"1625.0",\\
\quad \quad \quad "SID":"\texttt{<a\_156><b\_165><c\_10><d\_0>}"\},\\
\quad \quad \quad ...]\\
\quad \quad\},\\
\quad \quad ... \\
\quad ],\\
\quad "\textbf{current query}":\{"text":"customized badminton rackets", "time":"06/16 18:42:20", "location":"Xining"\}\\
\}\\
\midrule
\textbf{Output:}\\
<think>\\
The user is a male aged 31-40, indicating strong consumer capacity and a preference for high-end tech products, tending to choose powerful and reliable products... \\
The user exhibits a continuous interest in categories such as "badminton equipment" and "smartwatch accessories" based on their historical search behavior, especially with a high frequency of clicks on products from the "YONEX" brand. This indicates a higher level of brand loyalty and a tendency to choose established brands or best-selling items... \\
The user shows strong purchasing power and tends to choose items with high cost-effectiveness, clear functions, and reputable brands within budget... \\
For the current query ‘customized badminton rackets’, the predicted item brand is YONEX, as the user has a clear preference for this brand's products in the historical search... \\
Thus, the targeted product name should include keywords such as ‘new’, ‘professional’, ‘advanced’, and reflect high sales volume...\\
</think>\\
\texttt{<a\_60><b\_95><c\_95><d\_0>}\\
\bottomrule
\end{tabular}
\end{table*}
\vspace{-0.2in}
\end{document}